# Electromechanical Hysteresis in Phase Change Material $Sb_2S_3$

*Jack Kaman[1], Evan Musterman[2], Kyle P. Kelley[3], Neus Domingo-Marimon[3], Volkmar Dierolf[1], Himanshu Jain[1]\**

[1]Department of Materials Science and Engineering, Lehigh University, Bethlehem, PA 18015, United States

[2]National Synchrotron Light Source II, Brookhaven National Laboratory, Upton, NY 11973, United States

[3]Center for Nanophase Materials Sciences, Oak Ridge National Laboratory, Oak Ridge, TN, 37831, United States

*Email: hj00@lehigh.edu

Keywords: Piezoresponse Force Microscopy, Scanning Probe Microscopy, Phase Change Materials, Ferroelectric Semiconductors, Ion Dynamics

Abstract

Antimony sulfide is an emerging phase change material for optical and electrical memory and computation elements. It has additionally been reported as a ferroelectric, with recent evidence from hysteresis in piezoresponse force microscopy. Here, we complete a rigorous set of piezoresponse force microscopy experiments on a congruently crystallized $Sb_2S_3$ glass-ceramic, where piezoelectric coupling should be forbidden in glassy $Sb_2S_3$. We replicate previous results and reveal that the behavior is absent in glassy $Sb_2S_3$ but show that the response originates primarily from non-piezoelectric contributions to the signal caused by an applied voltage. This hysteretic behavior in piezoresponse force microscopy is quite similar to some electrochemically active non-ferroelectric oxides, but uniquely, it appears here with a very clear spatial contrast that is decoupled from surface topography. This shows that the electromechanical signal reflects bulk-like properties and reveals differences in electrical behavior of crystalline and amorphous phases of $Sb_2S_3$.

Antimony sulfide is a 'quasi-1D' semiconductor with potential applications in energy storage, photovoltaics, and as an optical and electrical phase change material, where the crystalline and amorphous phases have drastically different optical and electrical properties. These characteristics have enabled low-loss non-volatile tunable ring resonators, metastructures/metasurfaces, and electrical synaptic devices.[1–5] Several studies have also reported ferroelectricity in $Sb_2S_3$,[6–9] a property of technological interest because of its relevance for phase change applications and optoelectronics.

$Sb_2S_3$ crystallizes into an anisotropic structure featuring distorted, low-symmetry environments for Sb atoms. Qualitatively, the structure is similar to the prototypical layered orthorhombic GeS structure. However, Sb forms two types of distorted square pyramidal units with sulfur, which leads to corrugation along the layers, forming ribbonlike structures. This corrugated structure is similar to better studied ferroelectrics such as SbSI.[10–12] However, compared to other ferroelectric semiconductors, the ferroelectric properties of $Sb_2S_3$ are more sparsely discussed in literature.

Ferroelectric-like behavior has been reported in $Sb_2S_3$, such as weak polarization hysteresis, microwave-frequency second harmonic generation, and discontinuities in the temperature dependence of dielectric constant, DC conductivity, and ultrasound attenuation.[7–9,13–17] From these experiments, it is asserted that $Sb_2S_3$ assumes a slightly polar ground state below 420-450K, and a further increase in spontaneous polarization below 300-320K. Additionally, X-ray diffraction experiments [18,19] show very weak reflections indicating a crystalline space group with symmetry lower than Pnma below 420K and ~290K as well as changes in anisotropic thermal expansion coefficient corresponding to these phase transitions. While these reports support a polar structure, the most reported space group of $Sb_2S_3$ is orthorhombic and centrosymmetric Pnma[20–23], and other reports show no anomalies at these phase transition temperatures[23,24]. Given the discrepancies in literature, it is important to elucidate the relationships between measured behavior and underlying properties of $Sb_2S_3$. Here, we focus on the origin of the behavior of $Sb_2S_3$ as measured in Piezoresponse Force Microscopy (PFM), a common but nuanced characterization technique for nanoscale ferroelectric properties.[25]

PFM is a scanning probe technique where a cantilever is driven with a fast AC bias (~100s of kHz) and the displacement and/or flexure of the SPM tip is measured by the angle of a reflected laser incident on a photodetector. In a piezoelectric material, the amplitude and phase of this response will reflect the magnitude and sign of the out-of-plane piezoelectric constant. This fast AC bias can be superimposed on some larger DC bias program in to probe the material polarization as a function of electric field (fig. 1).[26,27] In analogy to macroscopic D-E hysteresis loops, PFM can be used to probe nanoscale switching behavior.[28,29] However, there are many mechanisms that can lead to an electromechanical coupling observed in PFM, especially in electrochemically active materials.[30–35] While ferroelectric hysteresis is characterized by stable polarization states and sharp switching thresholds (fig. 1b), hysteresis loops associated with non-piezoelectric phenomena are often unsaturated with characteristic wing-like distortions (fig. 1c) and can have a strong frequency or sweep rate dependence.[32] $Sb_2S_3$ nanowires yield unsaturated hysteresis loops,[25] indicating that the response may not be purely associated with piezoelectric coupling.

Without careful experimentation, it is often difficult to identify the mechanism for electromechanical coupling. PFM is often operated in a resonance-enhanced mode, where the electrical AC excitation is matched to the contact resonance of the cantilever, yielding an extremely high sensitivity to field-induced forces.[36] In resonance-enhanced PFM, finding a measurable signal in centrosymmetric or non-piezoelectric materials is common, and even more so after applying a large DC bias (as in experiments meant to probe ferroelectricity). Many such materials are electrochemically active, and notable examples include Y-stabilized $ZrO_2$, amorphous $HfO_2$, $CeO_2$, $NiO$, $Cu_2Mo_6S_8$, and many solid-state electrolytes.[30,31,35,37–40] It should be noted that there are numerous origins of coupling mechanisms, including electrostatic forces, surface-charge induced electrostriction, and chemical strain (Vegard strain).[32,41,42]

Here, in order to both clarify the physical properties of $Sb_2S_3$ and the origin of the PFM signal in an increasingly relevant class of semiconductive chalcogenide ferroelectrics/dielectrics, we leverage the glass-forming ability of congruently-crystallizing $Sb_2S_3$.[43] A cross-section of a partially crystallized $Sb_2S_3$ glass-ceramic was taken from a mm-sized sample fabricated via a melt-quench technique. Comparison of the behavior of the crystalline phase to the amorphous phase will show the effects of differences in electrical and electromechanical behavior, while environmental conditions and chemical conditions will be similar. First, we reproduce the hysteresis loops measured by Varghese et. al. in crystalline $Sb_2S_3$.[25]

**Band Excitation Polarization Switching (BEPS)**

A polarization switching PFM experiment applies a series of DC pulses in a triangular waveform at each pixel, while measuring the PFM response at 0 DC bias after each pulse to decrease the effects of electrostatics. In band excitation PFM, to mitigate effects associated with changes in resonant frequency, the cantilever is excited by a band of frequencies, and the phase and amplitude of the response are taken from a simple harmonic oscillator fit to the spectrum.[44] The PFM response, $D_{ac}$, is calculated as the amplitude times the cosine of the phase, which reflects the magnitude and sign of in-phase mechanical coupling to the applied voltage.

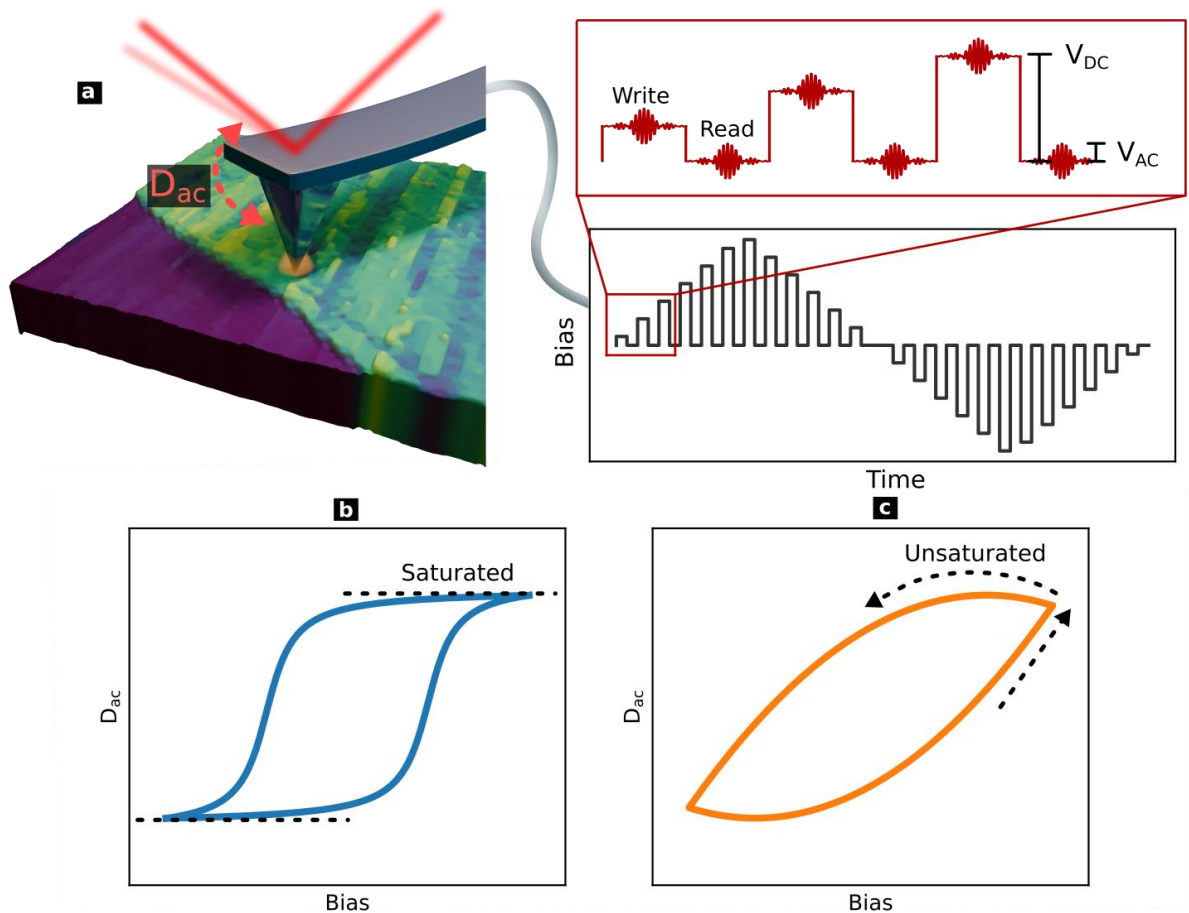

**Figure 1.** Schematic of a band excitation polarization switching (BEPS) experiment. (a) Scanning tip with typical BEPS bias program showing a small AC bias (a band of frequencies centered at

~100s of kHz) superimposed on a large DC (~1Hz) bias with a triangular envelope. (b,c) Qualitative illustrations of (b) a saturated hysteresis loop characteristic of ferroelectricity, and (c) an unsaturated hysteresis loop, typically associated with non-ferroelectric effects.

Figure 2 shows the results of the polarization switching-PFM experiment. Amplitude (fig. 2a) and phase (fig. 2b) show that there is no appreciable difference between the magnitude of the response in the crystal and glass before applying any large DC bias. The difference in phase appears to be associated with a difference in junction potential and is discussed in the following section. Figure 2d shows results from k-means clustering of PFM response dataset into two clusters. The hysteresis shown in figure 2d is a close match to the behavior found in $Sb_2S_3$ nanowires, with unsaturated response at high voltages and wing-like curvature that is mostly symmetric about 0 bias.[25] Qualitatively, the shape is most like those observed in NiO, $SrTiO_3$, and $TiO_2$ thin films, which are known to be electrochemically active and exhibit resistive switching and filament formation.[31,35] However, interestingly, glassy $Sb_2S_3$ yields a small, non-hysteretic response. A calculation of the loop area, shown in figure 2c, reveals a sharp contrast between the glassy and crystalline regions. However, before an applied DC bias, dual-ac resonance tracking (DART) shows a similar amplitude for glassy and crystalline $Sb_2S_3$ and no clear domain structures (See figure S1). This experiment shows that glassy and crystalline $Sb_2S_3$ behave similarly until a large bias is applied, after which crystalline $Sb_2S_3$ develops a large, hysteretic PFM response, while glassy $Sb_2S_3$ remains unchanged.

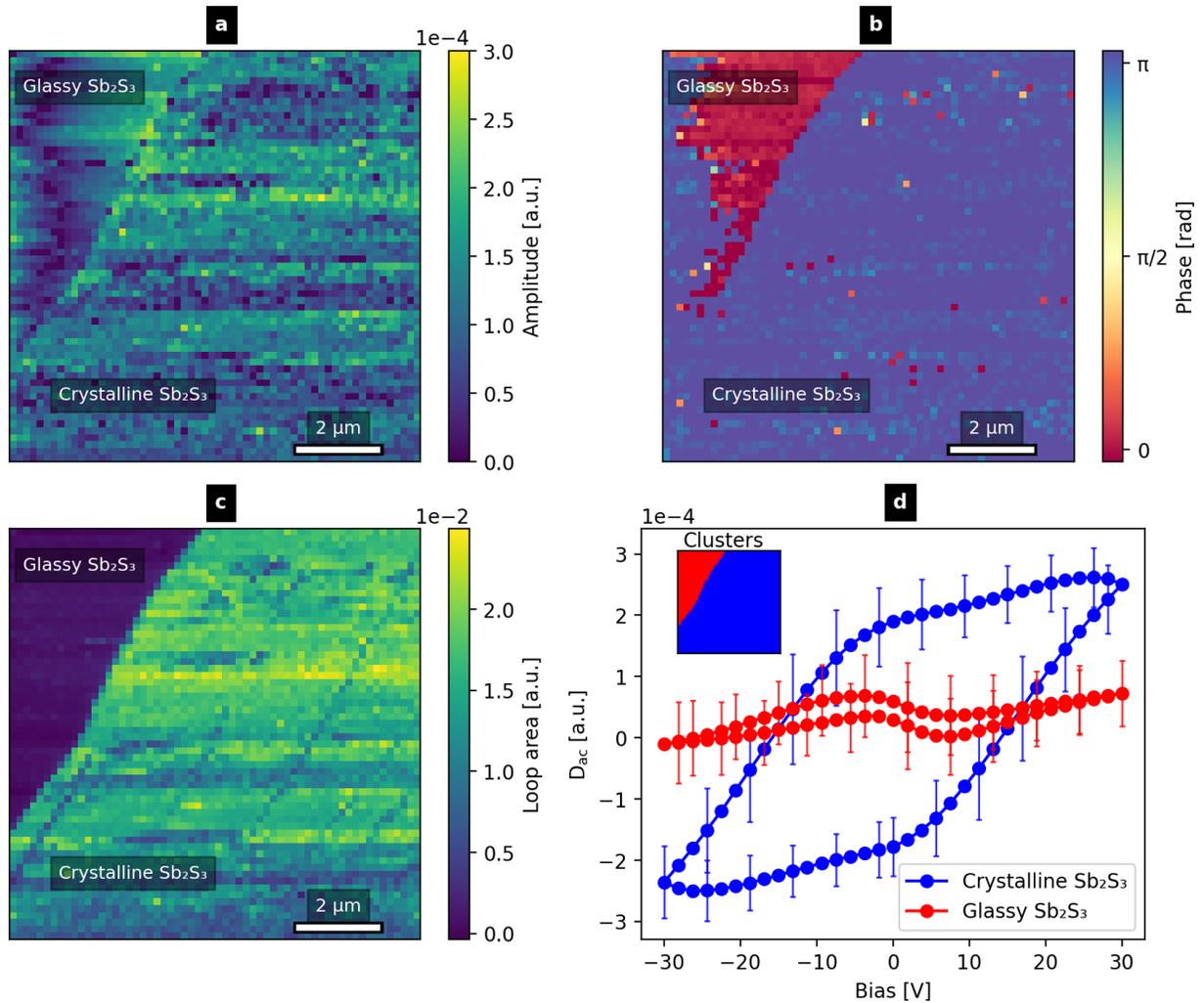

**Figure 2.** Band excitation polarization switching PFM map of $Sb_2S_3$ glass-ceramic, using a triangular DC bias of +/- 30V at 0.87 Hz. (a) Amplitude and (b) Phase of each point-by-point measurement before applying a DC bias. (c) Hysteresis loop area of PFM signal integrated over 4 DC bias cycles. (d) Averaged off-field PFM signal for 2 clusters (from k-means clustering) corresponding to crystalline $Sb_2S_3$ and glassy $Sb_2S_3$. Spatial image of clusters is shown in the inset.

Regardless of the coupling or charging mechanism, these results provide evidence that crystalline $Sb_2S_3$ yields a hysteretic surface polarization (or alternatively, surface charge). However, they do not imply that the surface charge originates from a switchable spontaneous polarization. To probe the stability of the surface polarization, we employ Kelvin Probe Force Microscopy (KPFM) to measure changes in (non-contact) surface potential.[45] After poling the surface of crystalline $Sb_2S_3$ with a scanning tip, KPFM shows only small changes in surface potential and reveals an absence of domain-like features (fig. S2). However, application of a ±30V triangle wave in a grid results in an increased surface potential and dimpled protrusions at each point (fig. S3). These features are correlated with a hysteretic response in BEPS, appearing exclusively in crystalline $Sb_2S_3$. Protrusions like these are common to find in electrochemically active materials after applying a large bias with a scanning probe, and in some cases, it has been attributed to electrochemical metallization, where a metal is precipitated on the surface of a metal oxide.[39,46] Although this implies ion migration that could be associated with a chemical (Vegard) strain, the sign of electromechanical coupling is determined by the charge and volume of a mobile ion (Vegard coefficient) and cannot switch sign under the influence of an applied bias unless there is more than one species causing a chemical strain. Thus, chemical strain is not a valid explanation for the hysteresis shown in figure 2. To further characterize the response, we use contact Kelvin Prove Force Microscopy (cKPFM).

**Contact Kelvin Probe Force Microscopy (cKPFM)**

In some materials such as relaxor ferroelectrics, ferroelectrics with wake-up effects, and ferroionics, lattice polarization is more complex and can be strongly coupled to ion migration, and some ferroelectrics can yield strangely shaped or strongly time-dependent hysteresis loops. A technique useful for differentiating ferroelectric and non-ferroelectric effects is contact Kelvin Probe Force Microscopy (cKPFM). A cKPFM voltage program contains a series of 'write' and 'read' steps, with the 'write' steps in a triangular waveform, similar to the polarization switching experiment. However, the response is 'read' at a range of nonzero biases to compare the relative strength of piezoelectric and non-piezoelectric signal. Mechanical signal generated by non-piezoelectric effects is often difficult to disentangle. Charged species (electrons/holes, ions) can accumulate to screen a tip bias, resulting in a concurrent surface charge, surface potential, and subsurface polarization. Although these can yield electromechanical coupling by different mechanisms, they are related by their origin of surface charging. Such bias-induced surface polarization can yield electromechanical coupling via electrostatic and/or electrostrictive coupling, with the non-piezoelectric electromechanical signal given by:

$$D_{ac}(V_{dc}) = k^{*-1} \cdot \frac{dC}{dz} \cdot V_{ac}\left(V_{dc} - V_{cpd}(V_{dc}, t)\right) + V_{ac} Q P_{surf}(V_{dc}, t)$$

**(1)**

where $D_{ac}$ is the AC displacement (electromechanical signal), $k^*$ is the effective tip-sample stiffness, $dC/dz$ is the derivative of capacitance with respect to (vertical) tip position, $V_{ac}$ is the excitation voltage amplitude, and $V_{cpd}$ is the potential difference measured in-contact.[37,47] For the electrostrictive term, Q is the effective electrostrictive coefficient, and $P_{surf}$ is the near-surface polarization. It can be seen from eq. 1 that, neglecting ferroelectric switching or dielectric nonlinearity, electromechanical response $D_{ac}$ should be a linear function of applied voltage $V_{dc}$. However, either a nonzero contact potential $V_{cpd}$ or a nonzero polarization $P_{surf}$ in a dielectric will result in a nonzero response at zero applied bias and appear as a left/right shift in cKPFM curves. Contrastingly, ferroelectrics, relaxors, and ferroionics should yield responses that are strongly nonlinear in voltage.[48–54] Measuring the response at several read biases ($V_{dc}$) will allow $V_{cpd}$ as a function of write bias to be estimated by finding the point where $D_{ac}$ is zero. Although electrostriction and electrostatic force are inseparable here, they both reflect the extent of induced polarization, and can be described by the x-intercept, $V_{Dac=0}$. Contact KPFM results for the same area as figure 2 are shown in figure 3.

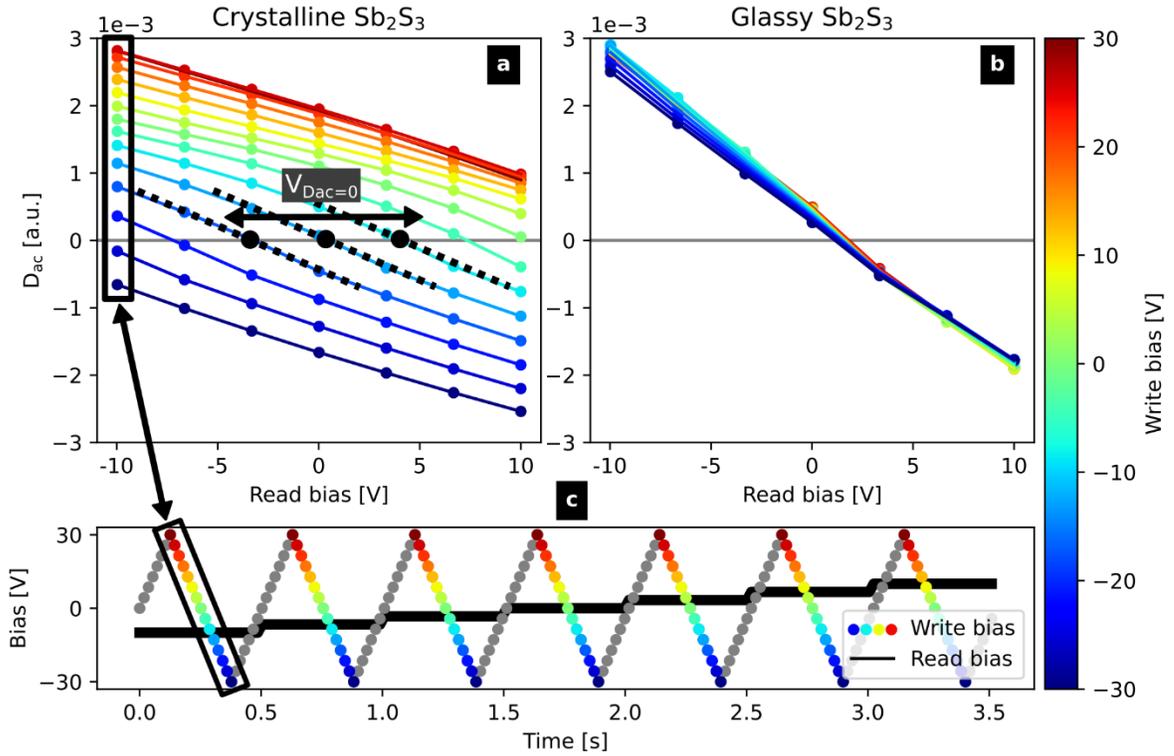

**Figure 3.** Contact Kelvin Probe Force Microscopy (cKPFM) experiment. Acquisition time for each point is 9ms. Response curves are colored for various write biases and plotted as a function of read bias. (a, b) Averages of clusters corresponding to (a) crystalline $Sb_2S_3$ and (b) glassy $Sb_2S_3$. (c) cKPFM voltage program. Points plotted in (a) and (b) are shown in color.

In glassy $Sb_2S_3$ (figure 3b), cKPFM reveals a linear electrostatic/electrostrictive contribution that is only very weakly dependent on the write voltage. In crystalline $Sb_2S_3$, on the other hand, the signal is linear with $V_{read}$ and has an x-intercept ($V_{Dac=0}$) that shifts greatly in voltage in response to the write bias. These measured values are much larger than the work function of $Sb_2S_3$ and should not be taken as a measurement of the contact potential difference $V_{cpd}$ but can nonetheless give insight into electrical behavior relevant to memory and computation devices.[55,56]

Because non - piezoelectric contributions to the PFM signal should be linear with this x-intercept, it is useful to compare the hysteresis of $V_{Dac=0}$ and $D_{ac}$.

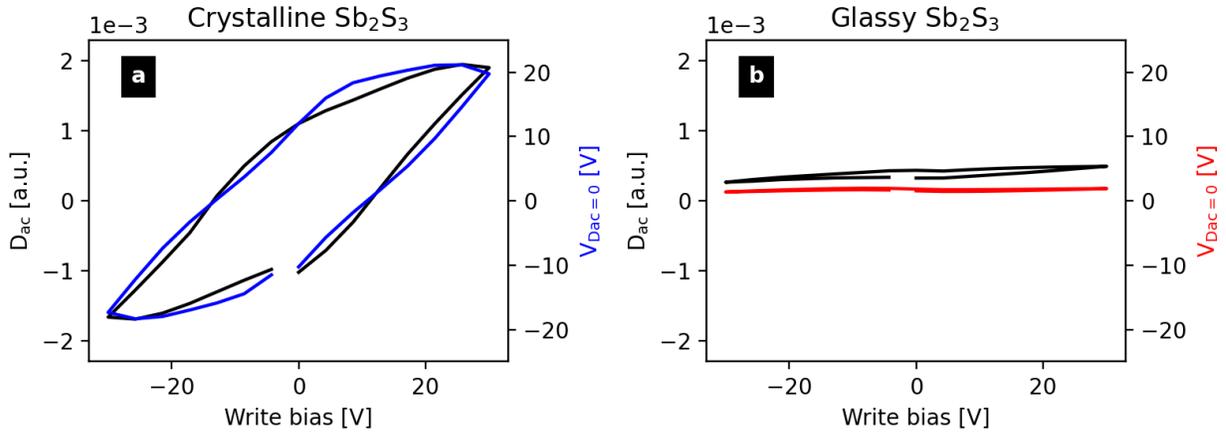

**Figure 4.** Response at $V_{read} = 0$ and bias offset $V_{Dac=0}$ calculated from cKPFM experiment. (a,b) Averages for clusters corresponding to (a) crystalline $Sb_2S_3$ and (b) glassy $Sb_2S_3$.

Figure 4(a) shows that the hysteresis of the x-intercept $V_{Dac=0}$ is quite similar to the hysteresis of the electromechanical signal $D_{ac}$, indicating that the main contribution to the measured 'PFM' signal in $Sb_2S_3$ is electrostatic forces or electrostriction associated with surface charging. Additionally, the non-saturating, rounded shape of the PFM signal near the peak writing bias is a general feature of lossy dielectrics.[57] Specifically, the surface charge (or surface potential) is dampened and delayed compared to the applied bias. Although it is not an exact analogy due to the complex read/write waveform, the dampened and phase-offset response corresponds to a lossy dielectric with strong dispersion around the frequency of the BEPS/cKPFM waveform (~2Hz). This surface charging, although not an electromechanical property of the material, has nonetheless been proven useful to understand local electrical behavior.[58–60] To further demonstrate how surface charge dynamics present themselves in PFM experiments and examine the electrical behavior of $Sb_2S_3$, we investigate the electromechanical response with relaxation spectroscopy.

**Electromechanical relaxation dynamics**

To probe the dynamics of the surface charging measured in cKPFM, we apply DC pulses of ±30V to polarize the sample surface while measuring the electromechanical response $D_{ac}$ versus time. Because the observed surface charging is on a timescale of a few Hz, we probe the dynamics on a timescale of seconds.

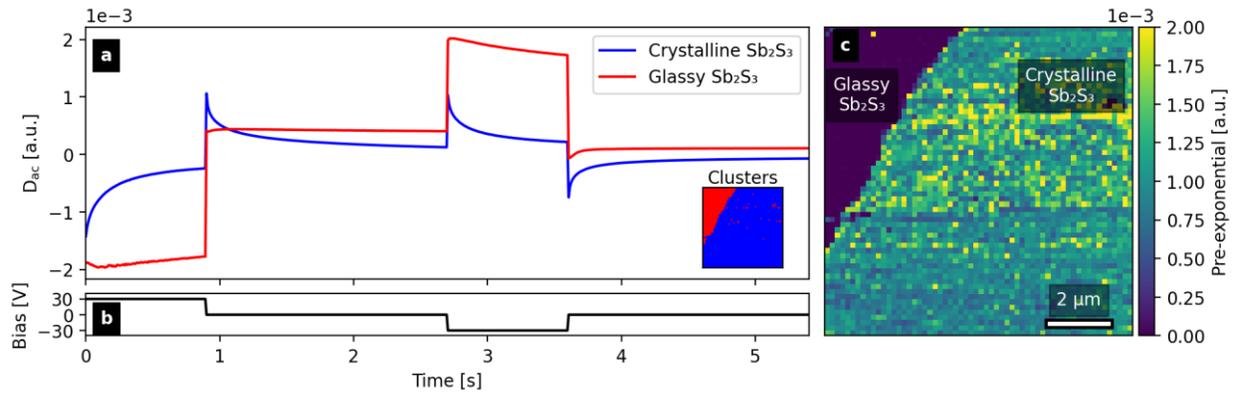

**Figure 5.** In-contact dynamic electromechanial experiment. (a) Response measured as a function of time for crystalline and glassy $Sb_2S_3$. (b) Applied bias as a function of time. (c) Pre-exponential factor of stretched exponential fit to transient $D_{ac}$ after +30V.

Here, dynamic electrostatic force experiments give a clear picture of the characteristics of $Sb_2S_3$. As expected from the cKPFM experiment, there is a fast/instantaneous electrostatic component of the response that has a negative coefficient (i.e., yielding a negative response under a positive applied bias). This is a feature of both glassy $Sb_2S_3$ and crystalline $Sb_2S_3$ as well as dielectrics in general[37], and since it appears globally in the scan area, it could be associated with tip-local and/or nonlocal (distributed) electrostatic forces. In crystalline $Sb_2S_3$, there is an additional local and time-dependent component of the electrostatic response that has a positive coefficient. In this case, surface charging slowly screens the applied $V_{dc}$. This surface charge decreases the on-field response by either screening the applied potential and decreasing the electrostatic contribution to $D_{ac}$, or alternatively, giving an electrostrictive coupling that opposes the electrostatic contribution. After the tip bias is removed, this leads to a nonzero force for a short time, explaining the hysteresis in BEPS measurements. The relaxation time is on the order of ~0.2s and would certainly lead to strong hysteresis when excited by a triangle wave of a similar period (~0.5s for cKPFM). Because the measured bias offset $V_{D_{ac}=0}$ is nearly linear with electromechanical response $D_{ac}$, the characteristics of the relaxation curves in a electromechanical relaxation experiment are a valid representation of the dynamic electrical properties of the tip-sample junction. This relatively long time constant allows us to characterize glassy $Sb_2S_3$ as an insulating dielectric and crystalline $Sb_2S_3$ as a (ionic and/or electronic) conductor.

Currently, the origin of the surface (or near-surface) charges is unclear. Surface charging is commonly attributed to long-range diffusion of charged defects, adsorbed species such as hydroxyl groups, or long-lived charge trapping at defect sites.[37] Here, the similar chemistry between the glassy and crystalline phase, combined with their starkly different behaviors, rule out surface adsorbed species as the origin of the signal, as such species would likely be present across the entire sample surface. This implies that the difference in behavior likely originates from the bulk. Indeed, crystalline $Sb_2S_3$ has low defect formation energies and crystallizes into a Van der Waals structure that could allow easy diffusion in some directions compared to glassy $Sb_2S_3$.[21,61–63] Additionally, many intrinsic defects have multiple charge states that could act as charge traps.[64,65] In a phase change material such as $Sb_2S_3$, the electronic structure, atomic environments, and defect properties are significantly different between the glassy and crystalline state,[63,66] and the origin of the difference in electrical behavior is difficult to discuss without further study.

Beyond characterizing the response as ferroelectric or non-ferroelectric, the ability to locally probe surface charge dynamics is valuable on its own. Strong PFM responses are common in electrochemically active materials, but the resulting changes are often small and correlated with topography, complicating interpretation.[67–69] Here, the electrical behaviors of glassy and crystalline $Sb_2S_3$ are uniquely distinct and decoupled from preexisting surface topography. Regardless of the origin of surface charging, strongly dynamic electrical behavior on timescales of milliseconds to seconds, especially with remnant behavior or non-exponential/heavy-tailed decay, has shown to be useful in devices for neuromorphic computing.[70,71] Further, here we show that the dynamic induced polarization is unique to crystalline $Sb_2S_3$, meaning that this dynamic charging behavior could be patterned by a laser/electron beam or set or reset by joule heating.[72–74] This could be used in a device with elements where dynamic behavior can be switched on (crystalline) and off (amorphous), which would bring the transient, fading behavior of dynamical memristors or memcapacitors to reconfigurable nonvolatile elements based on phase change materials.

In summary, these experiments allow us to characterize the PFM signal of crystalline $Sb_2S_3$ as an electrostatic or electrostrictive coupling originating from surface charging. More interestingly, we show that while glassy $Sb_2S_3$ appears to be a simply insulating dielectric, surface charging in crystalline $Sb_2S_3$ strongly screens an applied field and leads to a decaying remnant surface charge on a timescale of ~0.2s. We cannot exclude the possibility of piezoelectricity in $Sb_2S_3$ but conclude that any piezoelectric coupling in PFM is much smaller than the non-piezoelectric coupling observed here. This exemplifies the fact that PFM and related techniques require rigorous experimentation but can probe many phenomena that would otherwise be inaccessible, such as local electrical properties. The present experiments serve to clarify the origin of PFM signal in $Sb_2S_3$ and present new observations about the nanoscale electrical behavior of a technologically relevant phase change material.

Acknowledgments

This work was supported by an award from the US Department of Energy Office of Science, grant award no. DE-SC0005010. All scanning probe experiments were supported by the Center for Nanophase Materials Sciences (CNMS), which is a US Department of Energy, Office of Science User Facility at Oak Ridge National Laboratory.

Supporting Information for Electromechanical Hysteresis in Phase Change Material $Sb_2S_3$

*Jack Kaman, Evan Musterman, Kyle Kelley, Neus Domingo-Marimon, Volkmar Dierolf,*

*Himanshu Jain\**

**Sample Preparation**

A partially crystallized sample of Sb2S3 was fabricated by the melt-quench method by melting Sb2S3 powder (5N, Thermo Scientific) in an evacuated quartz capillary. The powder was melted at 650°C in a rocking furnace for ~12h. After removal, the sample was allowed to crystallize upon cooling for ~10 seconds and was subsequently water-quenched to stop growth. It was then sectioned and polished to a finish down to 0.05μm, yielding a smooth surface on a relatively thick (~0.5mm) sample of bulk Sb2S3 glass-ceramic.

**Figures**

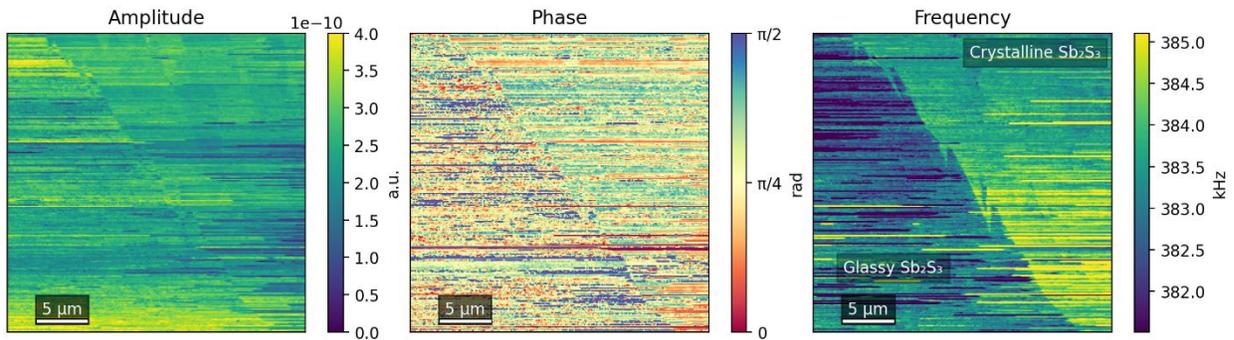

**Figure S1.** Amplitude, phase, and resonance frequency from Dual AC Resonance Tracking (DART) PFM.

Interestingly, in DART-PFM, the largest difference between glassy and crystalline $Sb_2S_3$ and is the resonance frequency, which scales with the tip-sample contact stiffness. Because the electrostatic part of the non-piezoelectric PFM signal scales inversely with contact stiffness (as given in eq. 1 in the main text), this could explain the shallow slope of cKPFM curves in crystalline $Sb_2S_3$.

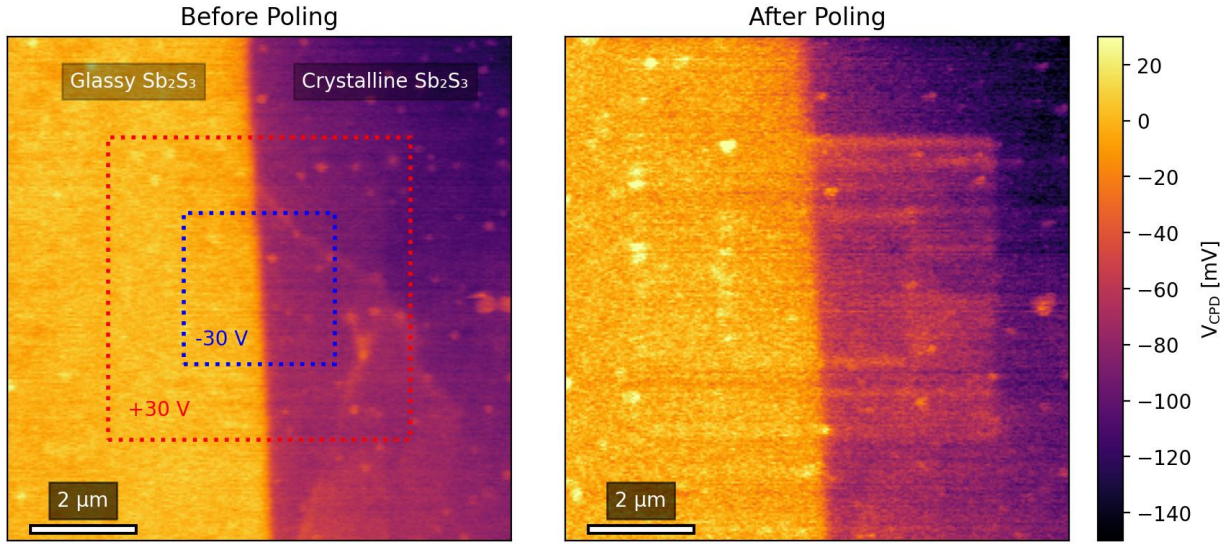

**Figure S2.** Surface potential from KPFM before and after a poling experiment where the tip was scanned at 12µm/s, applying +/- 30V.

KPFM reveals small (~100mV) differences in contact potential difference between glassy and crystalline $Sb_2S_3$. Although the magnitude is much smaller than the in-contact x-intercept in the cKPFM curves, the relative differences in potential are in agreement with the differences in phase shown in figure 1 and the horizontal offset in figure 2.

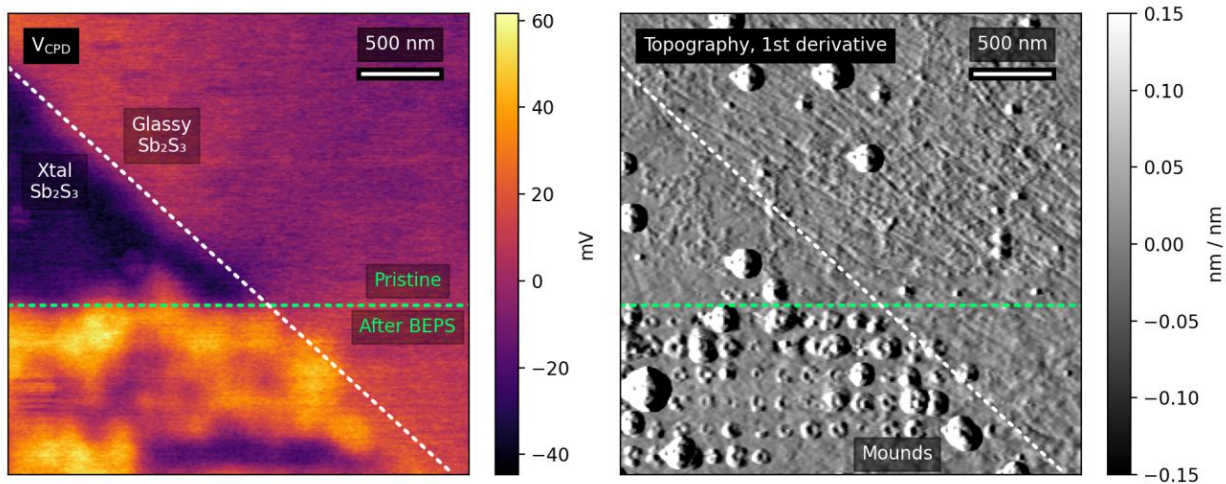

**Figure S3.** (left) Surface potential from KPFM after an experiment applying a triangular wave with a maximum of +/-30V at 2Hz. (right) Topography map after the experiment revealing protrusions on the surface of crystalline $Sb_2S_3$ at each point-by-point measurement. First derivative (horizontally) is shown for clarity.